\documentclass[a4paper,english,10pt,superscriptaddress,reprint,amsmath,amssymb,aps,prd,twocolumn,floatfix,notitlepage]{revtex4-2}
\usepackage{graphicx}
\usepackage{bm}
\usepackage{dcolumn}
\usepackage[colorlinks=true,citecolor=darkred,urlcolor=darkred, pdfborder={0 0 0}]{hyperref}
\usepackage{setspace}
\usepackage{caption}
\usepackage{subcaption} 
\usepackage{slashed}
\usepackage[normalem]{ulem}
\usepackage{float}
\usepackage[usenames,dvipsnames]{xcolor}  
\usepackage{microtype}
\usepackage{lipsum}
\usepackage{amsmath}
\usepackage{amssymb}
\usepackage{mathrsfs}
\usepackage{placeins}
\usepackage{lettrine}   
\input Eileen.fd        
\newcommand*\initfamily{\usefont{U}{Eileen}{xl}{n}}
\definecolor{darkred}{rgb}{0.6,0,0}
\definecolor{linkcolor}{rgb}{0,0,0.5}
\usepackage{orcidlink}
\graphicspath{{figs/}}
\usepackage[T1]{fontenc} 
\usepackage{tikz} 
\usetikzlibrary{arrows.meta}
\usetikzlibrary{decorations.text}
\usetikzlibrary{patterns.meta} 
\usepackage{tikz-layers} 
\usetikzlibrary{backgrounds, plotmarks}
\usepackage[compat=1.1.0]{tikz-feynhand}
\usepackage{tikz-feynman}
\tikzfeynmanset{compat=1.1.0}
\usepackage{feynmp}
\usepackage{tikzsymbols}
\usepackage{array}
\usepackage{pifont} 
\usepackage{mathrsfs} 

\definecolor{linkcolor}{rgb}{0,0,0.5}
\begin{document}
%
\preprint{2412.xxxx}
\title{\boldmath \color{BrickRed}Pathways to proton's stability via naturally small neutrino masses}
%
\author{Sin Kyu Kang \orcidlink{0000-0001-7508-3881}}
\email{skkang@seoultech.ac.kr}
\affiliation{School of Natural Science, \href{https://ror.org/00chfja07}{Seoul National University of Science and Technology}, \\232 Gongneung-ro, Nowon-gu, Seoul, 01811, Korea}
\author{Oleg Popov \orcidlink{0000-0002-0249-8493}}
\email{opopo001@ucr.edu (corresponding author)}
\affiliation{Faculty of Physics and Mathematics and Faculty of Biology, \href{https://ror.org/02q963474}{Shenzhen MSU-BIT University},\\ 1, International University Park Road, Shenzhen 518172, China}
\date{\today}
%
\begin{abstract}
In the present work, the connection between the smallness of the neutrino masses and the stability of the proton is studied. We analyze this connection from different perspectives: the smallness of neutrino mass and the proton stability originate from the same source, small neutrino masses lead to a long lived proton, and the smallness of the proton decay width as a cause of the naturally small neutrino masses. All the schemes are studied in detail and UV realizations are given. We discuss advantages of each scheme and outline further investigation directions.
\end{abstract}
\maketitle 
\section{Introduction}
\label{sec:intro}
\lettrine[lines=4,findent=-1.3cm]{\normalfont\initfamily \fontsize{17mm}{10mm}\selectfont S \normalfont\initfamily}{ }tandard model (SM) has been established by numerous experimental tests, yet evidences on neutrino mass via neutrino oscillation, the mystery of dark matter, and baryon asymmetry of the universe call for beyond standard model (BSM) physics. 

While the proton is considered a stable particle in the renormalizable SM due to the conservation of baryon number, numerous BSM models predict proton decay mediated by new particles. Grand Unified Theories (GUTs) serve as representative examples, suggesting proton decay with a lifetime closely tied to the grand unification scale.
Thus, proton decay is one of the key predictions of the various GUT where the proton has a half-life of about $10^{31}-10^{36}$ years and decays into a positron and a neutral pion that itself immediately decays into two gamma ray photons~\cite{Georgi:1972cj,Georgi:1974yf}.
The experiments dictate the proton mean lifetime to be more than $10^{30}-10^{34}$ years
~\cite{ParticleDataGroup:2022pth,Senjanovic:2009kr}.

On the other hand, massive neutrinos demand BSM physics. Neutrino masses must be exceptionally small, prompting us to explore alternative mechanisms for imparting mass to charged fermions, distinct from the conventional approaches. The scale of BSM physics may be the origin of smallness of neutrino masses. 
The most popular mechanism to generate tiny neutrino masses is seesaw mechanism.
The scotogenic scenario is a good alternative to the seesaw mechanism, where tiny neutrino masses can be obtained by radiative 1-loop corrections and a dark matter candidate is naturally accommodated \cite{Ma:2006km}. This scenario is also applied to the proton decay occurred at loop level as studied in ~\cite{Gu:2016ghu,Helo:2019yqp}.

In the context of GUT, both proton decay and massive neutrinos can be simultaneously accommodated.
In this work, we propose new idea on the link between proton decay and massive neutrinos. 
Contrary to GUT, both proton decay and neutrino masses are generated via only loops. Both suppression of the proton  decay rates and small neutrino masses have common origin attributed to the same mediators inside loops.
Hence, smallness of neutrino masses can lead to the suppression of the proton decay rate, and vice verse.
Also, it is possible to explain both simultaneously through the same mediators.
By product, we show that some models contain dark matter candidates like in the scotogenic model for massive neutrinos \cite{Ma:2006km} as well as proton decay ~\cite{Gu:2016ghu,Helo:2019yqp}.
%
%
%
%
\section{Concept of connection between neutrino mass and proton decay}
\label{sec:connection_concept}
Traditionally, generation of neutrino mass and proton decay have been considered as separate problems  in the literature. Neutrino mass generation is usually achieved with minimal BSM extension. In further cases origin of neutrino masses is considered in conjunction with dark matter, cosmological applications, and other phenomena. On the other hand, proton decay or life time dominantly appears in the GUTs, where it is a natural consequence. With all this said, we see that these two problems in high energy physics appear largely disconnected.

This is the place where we put forward the idea that these two problems can have one solution and show how it can be done. The proposal is that the origin of neutrino masses is connected with the life time of the proton. To put it in another way, the naturality of the smallness of the neutrino masses and the longevity
of the proton are connected or stem from the same source.

Three, conceptually distinct, connections are possible and are explained further. In Sec.~\ref{sec:connection_realization}, the UV complete models of aforementioned concepts are described and some further details are included.

First connection represents the common origin of neutrino mass and proton decay operator. Schematically it can be viewed as in 
Fig.~{\color{red}1a} of the supplementary materials. One source generates two effects: tiny neutrino masses and long proton lifetime, which are correlated. In the example UV model given in the next section, the source is dark sector (DS) that leads to both radiative neutrino mass origin and radiative proton decay operator. As a result, and will be shown in Sec.~\ref{sec:com_origin}, DM mass, neutrino mass, and proton decay are correlated. In the case if the source vanishes, both the neutrino mass and the proton decay width approach zero. In other words, in the limit of massless neutrinos, the proton becomes stable.

The second connection, shown in 
Fig.~{\color{red}1b} of the supplementary materials, demonstrates a scenario where proton decay serves as the source and the neutrino mass emerges as the result. As a consequence, radiative neutrino mass is generated from the proton decay operator, giving rise to a correlation between the two. In the case of stable proton, \emph{i.e.} when the proton decay vanishes, neutrinos become massless.

The third connection describes how the proton decay may be generated from tiny neutrino masses. The case is shown schematically in 
Fig.~{\color{red}1c} of the supplementary materials, where the source is the neutrino mass and the result is the proton decay that vanishes in the limit of massless neutrinos.

All three types of connections between neutrino mass and proton decay have their own advantages and require distinct minimal BSM content. The concept of connection between neutrino mass and proton decay is schematically represented in Fig.~{\color{red}1}
of the supplementary materials.
%
\section{Realizations of connection between neutrino mass and proton decay}
\label{sec:connection_realization}
Outlined below are the three UV complete models that connection neutrino mass and proton decay origins according to the casualty directions depicted in Fig.~{\color{red}1}
of the supplementary materials.
%
\subsection{Common origin to smallness of neutrino mass and proton stability}
\label{sec:com_origin}
%
%
In this, extended scotogenic model, scenario both the smallness of the proton decay width and the smallness of the neutrino mass are naturally achieved through the existence of the dark matter $N$. The fermionic dark matter $N$ mediates both the proton decay operator and the Weinberg dim-5 neutrino mass operator radiatively at one-loop order, which in turn make both of the resultant effects naturally small. Present scenario demonstrates how the smallness of the neutrino mass and the large lifetime of the proton can be simultaneously obtained and correlated via a common source, the dark sector in this model. The vector-like heavy EW singlet quark $D\sim\left(\pmb{3},\pmb{1},-1/3,-\right)$ and $\Tilde{R}_{2D}\sim\left(\pmb{3},\pmb{2},1/6,-\right)$ leptoquark under SM and $\mathbb{Z}_2$ symmetries are added in addition to the canonical scotogenic fields.
Tab.~{\color{red}I} of the supplementary materials represents the field content of the extended scotogenic model.

The Yukawa Lagrangian and scalar potential for the model fields content are as follows
\begin{subequations}
\label{eq:lag}
\begin{align}
    \label{eq:lag_yuk}
    -\mathcal{L}^{\text{Yuk}}_{\text{BSM}} &= Y_1 \Bar{D} L \Tilde{R}_{2D} + Y_2 Q N \Tilde{R}_{2D}^\dagger + Y_3 Q D \Tilde{R}_{2D}  \\
     &+ Y_4 L N \eta^\dagger + Y_5 Q \Bar{D} \eta 
     \nonumber \\
    &+ M_N N N + M_D D \Bar{D} + \text{h.c.}, \nonumber \\
    \label{eq:lag_pot}
    V &= m_H^2 \left(H^\dagger H\right) + m_\eta^2 \left(\eta^\dagger \eta\right) + m_{R}^2 \left(\Tilde{R}_{2D}^\dagger \Tilde{R}_{2D}\right) \nonumber \\
    &+ \lambda_H \left(H^\dagger H\right)^2 + \lambda_\eta\left(\eta^\dagger \eta\right)^2 + \lambda_{R} \left(\Tilde{R}_{2D}^\dagger \Tilde{R}_{2D}\right)^2 \nonumber \\
    &+ \lambda_{R}^\prime \left(\Tilde{R}_{2D}^\dagger \Tilde{R}_{2D} \Tilde{R}_{2D}^\dagger \Tilde{R}_{2D}\right) + \lambda_{H\eta} \left(H^\dagger H\right)\left(\eta^\dagger \eta\right) \nonumber \\
    &+ \lambda_{H\eta}^\prime \left(H^\dagger \eta\right)\left(\eta^\dagger H\right) + \lambda_{HR} \left(H^\dagger H\right)\left(\Tilde{R}_{2D}^\dagger \Tilde{R}_{2D}\right) \nonumber \\
    &+ \lambda_{HR}^\prime \left(H^\dagger \Tilde{R}_{2D}\right)\left(\Tilde{R}_{2D}^\dagger H\right) \nonumber \\
    &+ \lambda_{\eta R} \left(\eta^\dagger \eta\right) \left(\Tilde{R}_{2D}^\dagger \Tilde{R}_{2D}\right) + \lambda_{\eta R}^\prime \left(\eta^\dagger \Tilde{R}_{2D}\right) \left(\Tilde{R}_{2D}^\dagger \eta\right) \nonumber \\
    &+\left( \lambda H H \eta \eta + \lambda_{3R} \Tilde{R}_{2D} \Tilde{R}_{2D} \Tilde{R}_{2D} \eta + \text{h.c.}\right)
\end{align}
\end{subequations}

Feynman diagrams generated via a common dark sector of the correlated radiative proton decay and neutrino mass in this model are sketched in Fig.~\ref{fig:sce_1_p_decay} (radiative proton decay) and in Fig.~{\color{red}2}
 (radiative neutrino mass) of the supplementary materials' Sec.~A.
\begin{figure}[ht]
\centering
    \centering
    \includegraphics[width=0.38\textwidth]{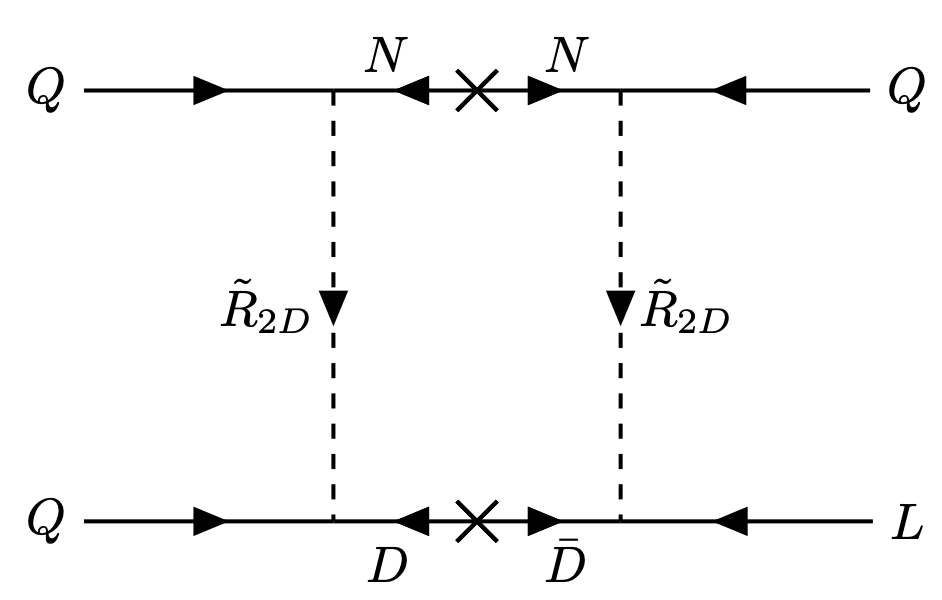}
\caption{Proton decay Feynman diagram of the extended scotogenic model.}
\label{fig:sce_1_p_decay}
\end{figure}

The proton decay width generated by Feynman diagram in Fig.~\ref{fig:sce_1_p_decay} is given as
\begin{widetext}
\begin{align}
\label{eq:p_decay}
        \Gamma \left(p\rightarrow e^+ \pi^{0},\nu_\alpha^\dagger \pi^{+} \right) &= \frac{m_p}{512\pi^3} \left(1-\frac{m_\pi^2}{m_p^2}\right)^2 \left| Y_2^{1a} \frac{M_N^{ab}}{m_{R}} Y_2^{b1} Y_3^{1c} \frac{M_D^{cd}}{m_R} Y_1^{d\alpha} \frac{W_0^{l}\left(p\rightarrow e^+ \pi^{0},\nu^\dagger \pi^{+}\right)}{m_R^2} \right. \\
        &\times\left. \left[ F_0 (x_N, x_D, y) + F_1 (x_N, x_D, y) \frac{q^2}{m_R^2} \right] \right|^2 , \nonumber
\end{align}
\end{widetext}
where for $p \rightarrow e^+ \pi^0$,  $\alpha = e$, since the constraints on a $p \rightarrow e^+ \pi^0$ are more stringent compared to $p \rightarrow \mu^+ \pi^0$. On the contrary, for neutrinos situation differs from $p \rightarrow e^+ \pi^0$. So, for $p \rightarrow \nu_\alpha^\dagger \pi^+$, $\alpha = e,\mu,\tau$, since the neutrino escapes and can be any flavor, whereas experimental constraint applies to $p \rightarrow \nu \pi^+$. Loop functions, Kallen $\lambda$, and other relevant definitions are given in the supplementary materials. Here, $m_p$ is a proton mass, $m_\pi$ is a pion mass, $q$ is the momentum exchanged between quarks in the decay process, whereas $a,b,c,d$ are the flavor indices.

Proton decay matrix elements and exchanged momentum between the proton's quarks during decay process are obtained from the lattice calculations~\cite{Aoki:2017puj} and are given below

\begin{subequations}
    \label{eq:p_decay_mat_elem}
    \begin{align}
        \label{eq:p_decay_mat_elem_1}
        W_0^e (p \rightarrow e^+ \pi^0) &= \left\langle \pi^0 \right| (ud)_L u_L \left|p\right\rangle \nonumber \\
        &= 0.134(5)(16)~\text{GeV}^2\text{ \cite{Aoki:2017puj}}, \\
        \label{eq:p_decay_mat_elem_2}
        W_0^\nu (p \rightarrow \nu^\dagger \pi^+) &= \left\langle \pi^+ \right| (du)_L d_L \left|p\right\rangle \nonumber \\
        &= 0.189(6)(22)~\text{GeV}^2\text{ \cite{Aoki:2017puj}}, \\
        \label{eq:p_decay_mat_elem_q}
        q^2 &= 0.2~\text{GeV}^2\text{ \cite{Aoki:2017puj}}.
    \end{align}
\end{subequations}

The $W_1^{l}$ for $l=e^+$ or $l=\nu_\alpha^\dagger$ are negligible compared to $W_0^l$~\cite{Aoki:2017puj}.

Scotogenic neutrino mass~\cite{Ma:2006km} generated via a Feynman diagram in Fig.~{\color{red}2}
of the supplementary materials' Sec.~A is given as
\begin{subequations}
\label{eq:sce_1_mnu}
\begin{align}
    &m_\nu^{ij} = \frac{1}{32\pi^2} Y_4^{ik} M_{N_k} Y_4^{kj} \left[ f\left(\frac{m_{\eta_R}^2}{M_{N_k}^2}\right) - f\left(\frac{m_{\eta_I}^2}{M_{N_k}^2}\right) \right], \\
    &m_\nu^{ij} \approx -\frac{1}{16\pi^2} \frac{\lambda v^2}{M_{\eta}^2} Y_4^{ik} M_{N_k} Y_4^{kj},
\end{align}
\end{subequations}

where $f(x) = \frac{x}{1-x} \ln x$ and $M_{\eta}^2 = \frac{ m_{\eta_R}^2 + m_{\eta_I}^2 }{2}$.

Whereas this construction requires more BSM fields, compared to the models studied in the following sections of the manuscript, one clear advantage of this scenario is the connection between smallness of the neutrino mass, longevity of the proton, and the existence of dark matter. Detailed study of this construction and its consequences can be found in~\cite{Nomura:2024zca}.
%
\subsection{Smallness of neutrino mass via proton stability}
\label{sec:mnu_to_p_decay}
In this realization we achieve a naturally suppressed neutrino masses through the proton decay operator, which will lead to the dependence of the earlier on the later and if the proton's longevity is large then, as a consequence, the neutrino masses must be tiny. The model introduced is composed of SM particles and an amendment of $S_3$ and $S_3^\prime$ leptoquarks. Furthermore, the SM symmetries are appended by global continuous abelian $U(1)_F$ symmetry, where $F=3B-L$, $B$, and $L$ are the fermion, baryon, and lepton numbers of the particles in the model, respectively. The field content of the model is given in 
Tab.~{\color{red}II} of the supplementary materials. The two leptoquarks introduced in the model transform identically under SM symmetries and differ only by $U(1)_F$ charges. This is done to maintain the $U(1)_F$ symmetry in tact and break it explicitly and softly, only by the dim-2 term in the Lagrangian.
\begin{figure*}
\centering
    \begin{subfigure}{0.45\textwidth}
    \centering
    \includegraphics[width=1.0\textwidth]{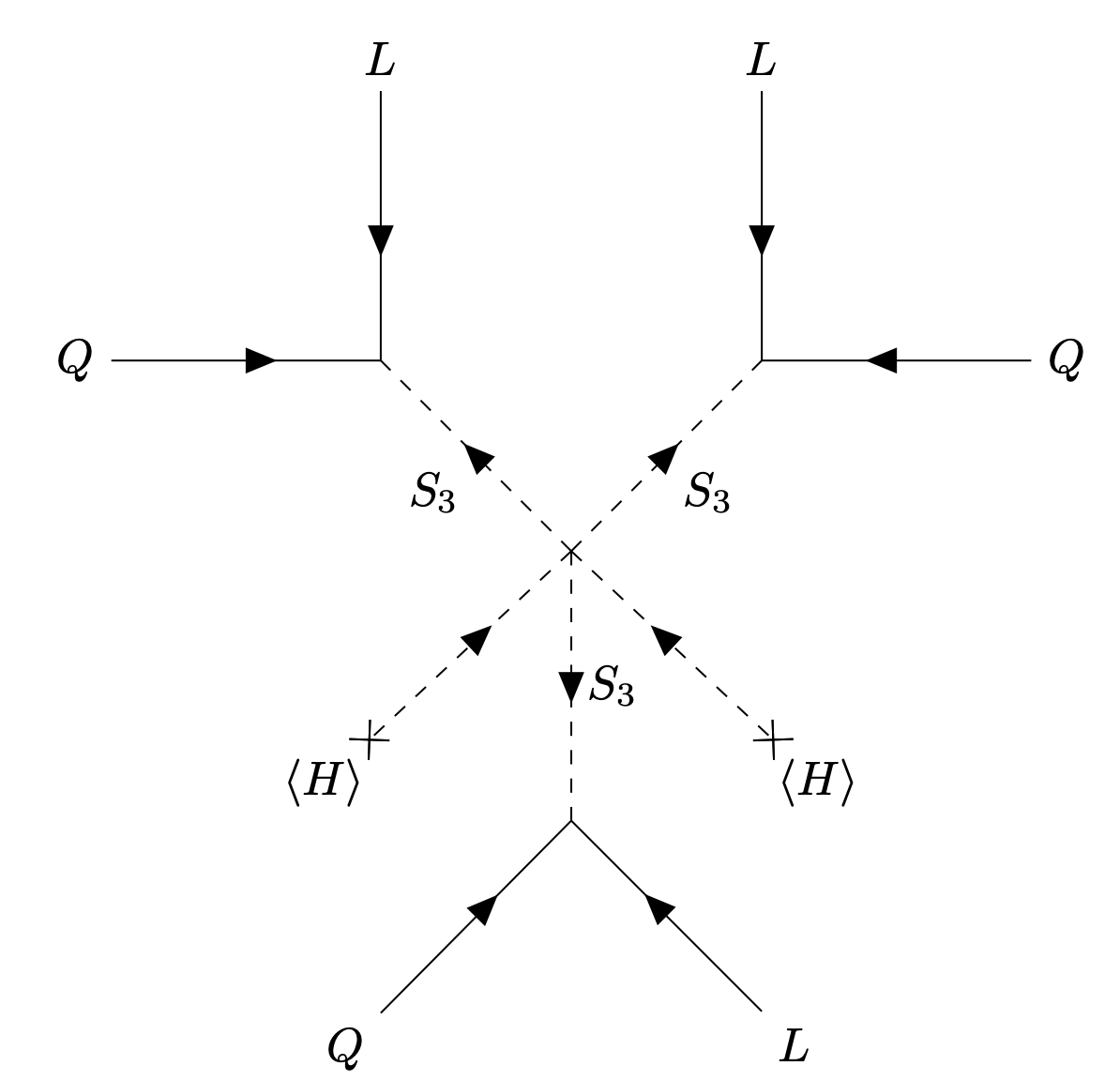}
    \caption{Operator contributing to proton decay ($p\rightarrow e^+\nu^\dagger\nu^\dagger$) with $\Delta \left(3B-L\right) = 0$.}
    \label{fig:p_decay_3b-l_o1_sce_2}
    \end{subfigure}
    \begin{subfigure}{0.45\textwidth}
    \centering
    \includegraphics[width=1.0\textwidth]{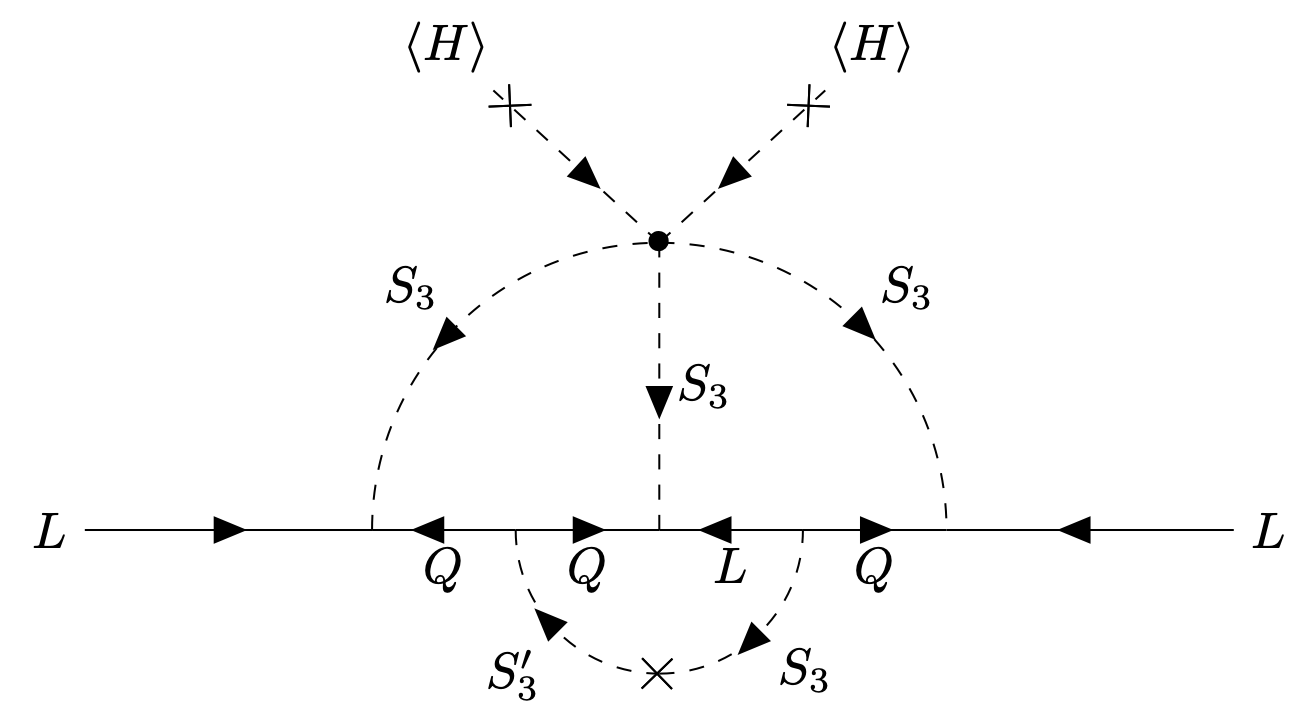}
    \caption{3-loop Weinberg operator of the neutrino mass.}
    \label{fig:3l_weinberg_sce_2}
    \end{subfigure}
    \caption{Framework and Feynman diagrams of 3-body proton decay and its induced radiative neutrino masses.}
    \label{fig:sce_2_dia_1}
\end{figure*}

The Yukawa Lagrangian and scalar potential for the model fields are as follows
\begin{subequations}
\label{eq:lag2}
\begin{align}
    \label{eq:lag_yuk2}
    -\mathcal{L}^{\text{Yuk}}_{\text{BSM}} &= Y_1^{\prime} Q L S_{3} + Y_2^{\prime} Q Q S_{3}^{\prime \dagger}  + \text{h.c.} \\
    \label{eq:lag_pot2}
    V &= m_H^2 \left(H^\dagger H\right) + m_{S_3}^2 \left(S_3^\dagger S_3\right) +  m_{S^{\prime}_3}^2 \left(S_3^{\prime \dagger} S_3^{\prime}\right) \nonumber \\
    &+ \lambda_H \left(H^\dagger H\right)^2 + \lambda_{S_3}\left(S_3^\dagger S_3\right)^2 + \lambda_{S^{\prime}_3} \left(S_3^{\prime \dagger} S_3^{\prime}\right)^2 \nonumber \\
    &+ \lambda_{HS_3} \left(H^\dagger H\right)\left(S_3^\dagger S_3\right) + \lambda_{HS_3} \left(H^\dagger H\right)\left(S_3^{\prime \dagger} S_3^{\prime}\right)\nonumber \\
    &+ \lambda_{S_3S^{\prime}_3} \left(S_3^\dagger S_3\right)\left(S_3^{\prime \dagger} S_3^{\prime}\right) + \lambda_{HS_3}^{\prime} \left(H^\dagger S_3\right)\left(S_3^\dagger H \right) \nonumber \\
    & + \lambda_{HS_3^{\prime}}^{\prime} \left(H^\dagger S_3^{\prime}\right)\left(S_3^{\prime \dagger} H \right) + \lambda^{\prime}_{S_3S^{\prime}_3}\left(S^{\dagger}_3 S^{\prime}_3\right)\left(S^{\prime \dagger}_3 S_3\right) \nonumber \\
    &+ \left( \frac{\lambda_{HSS}}{M}\left(H^{\dagger} H^{\dagger} S_3 S_3 S_3\right)+\text{H.c.}\right).
\end{align}
\end{subequations}
Here, we have included $D$-5 term in $V$.

Proton decay will proceed via several channels in the present scenario. The $F$ number conserving operator that contributes to the three-body proton decay is generated by the Feynman diagram given in Fig.~\ref{fig:p_decay_3b-l_o1_sce_2}. Therefore, proton decay is suppressed by the leptoquark mass and phase space of the final states. There is also a proton decay generated through the soft $F$ number violating term $m_3^2 S_3^\dagger S_3^\prime + $ h.c. The corresponding proton decay channel is illustrated in Fig.~{\color{red}3a}
of the supplementary materials' Sec.~B. This results in two-body proton decays by means of $U(1)_F$ violation and is suppressed by the $m_3$ parameter alone. The relative domination of the two-body decay channels depends only on $S_3$ and $S_3^\prime$ leptoquarks' masses and their mixing parameter $m_3$. The correlation between neutrino mass and proton decay, including the relative contribution of the two channels to the total proton decay width, will be studied elsewhere. The tiny radiative neutrino masses are generated by the Weinberg dim-5 neutrino mass operator at the three-loop order, as shown in Fig.~\ref{fig:3l_weinberg_sce_2}. The effective version of this Feynman diagram and direct correlation between neutrino mass and proton decay amplitude can be seen from Fig.~{\color{red}3b} of the supplementary materials' Sec.~B.

The three-body proton decay width, obtained from the Feynman diagram in Fig.~\ref{fig:p_decay_3b-l_o1_sce_2}, is given as
\begin{subequations}
    \label{eq:s3_p_decay}
    \begin{align}
        \Gamma \left( p\rightarrow e^+\nu^\dagger\nu^\dagger \right) &= \frac{m_p}{\left( 8\pi \right)^3} \left| \frac{\lambda_{HSS} v^2}{M m_S} \frac{\left(Y_1^{\prime}\right)^{1\alpha} \left(Y_1^{\prime}\right)^{1\beta} \left(Y_1^{\prime}\right)^{11}}{\left[ 1 - q^2 / m_S^2 \right]^3} \right. \nonumber \\
        &\left.\times\frac{W_0^{l}\left(p\rightarrow e^+\nu^\dagger\nu^\dagger \right)}{m_S^5} \right|^2,
    \end{align}
\end{subequations}
where $\alpha, \beta$ are the final-state neutrino flavor indices, $q$ is the 4-momentum exchanged between the quarks of the proton, $m_S$ is the $S_3$ leptoquark mass, $m_p$ is the proton mass, $v$ SM Higgs VEV, and $W_0^l$ is the proton decay matrix element of the corresponding channel that is to be computed on the lattice.

To summarize the model, neutrino masses arise from the proton decay operator by the minimum BSM extension, which requires two copies of the $S_3\sim(\bar{3},3,1/3)$ leptoquark and $U(1)_F$ global symmetry beyond the SM symmetries. In the limit of the stable proton, \emph{i.e.} the limit $m_3\rightarrow 0$ and $\lambda_{HSS}\rightarrow 0$, the neutrino masses vanish identically. As a result of this construction, naturally tiny neutrino masses are generated by means of proton's longevity.
%
\subsection{Proton stability via smallness of neutrino mass}
\label{sec:p_decay_to_mnu}
\begin{figure}[t]
\centering
    \includegraphics[width=0.45\textwidth]{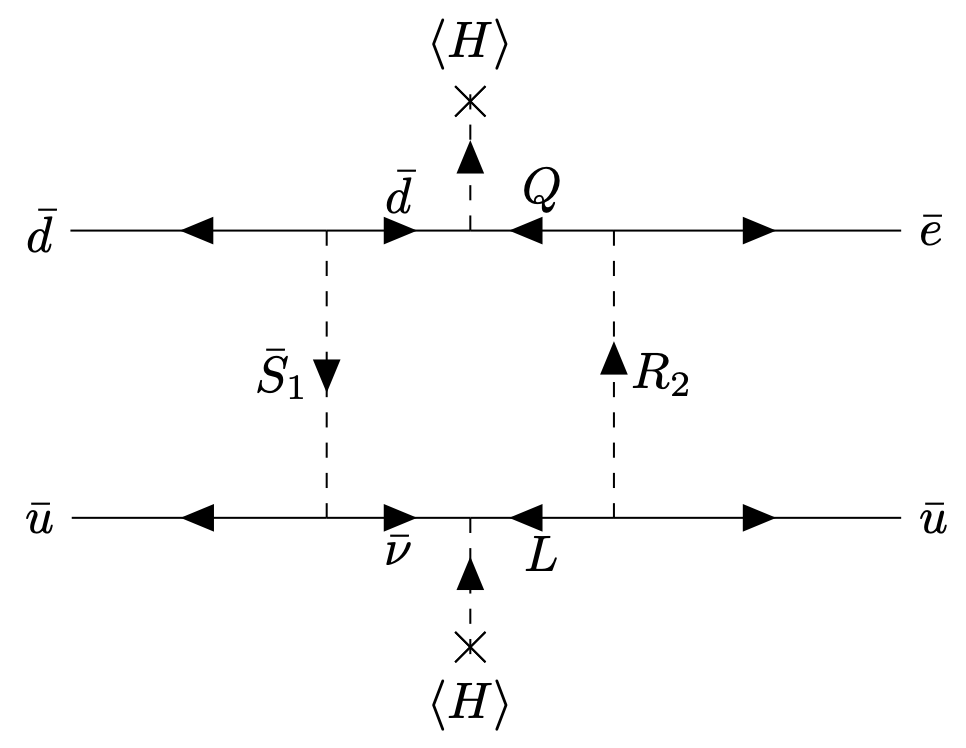}
    \caption{Feynman diagram contributing to proton decay in the neutrino mass mediated proton decay framework.}
    \label{fig:sce_3_p_decay}
\end{figure}

The field content of the minimal model is given in 
Tab.~{\color{red}III} of the supplementary materials. The model is minimal in the sense that it requires no BSM symmetries, no dark sector, and it is realized at one-loop order(Fig.~\ref{fig:sce_3_p_decay}) and tree level via neutrino mass mixing (Fig.~\ref{fig:s1_p_decay_2}). In order to avoid unwanted proton decay channels (Fig.~{\color{red}4}
of the supplementary materials' Sec.~C) that are independent of the neutrino mass, the neutrinos are made Majorana with a sterile neutrinos' mass being of the order of the seesaw scale ($\sim\mathcal{O}\left(10^{12}\right)$~GeV). Tiny neutrino masses are generated by the seesaw-I mechanism, the Feynman diagram of which is given in Fig.~{\color{red}5} of the supplementary materials' Sec.~C. The seesaw mass is given by $m_\nu \approx - \left( Y_5^{\prime\prime} v \right)^2 / M_\nu$, where $v$ is the EW Higgs doublet VEV and $M_\nu$ is the Majorana mass of the heavy neutrinos.
%
%

This suffices to make leading order proton decay proportional to the neutrino mass, owning to the unique properties of the $\Bar{S}_1\sim\left(\pmb{\Bar{3}},\pmb{1},-2/3\right)$ leptoquark, it only couples to the $\Bar{\nu}$ of SM leptons. Therefore, all proton decay channels, mediated by $\Bar{S}_1$, lead to proton decays with sterile neutrino in the final state, \emph{i.e.} $p\rightarrow \Bar{\nu} X^+$ where $X$ is light charged meson, and in the case of Majorana neutrinos with sterile neutrinos being of the seesaw scale, all these channels are kinematically forbidden. $R_2\sim\left(\pmb{\Bar{3}},\pmb{2},7/6\right)$ leptoquark is a non-chiral leptoquark of the genuine kind~\cite{Dorsner:2016wpm} (it only has leptoquark couplings to the SM fermions and no diquark couplings), therefore it does not contribute to the proton decay at the leading order.~\footnote{The leading order contribution of $R_2$ leptoquark to the proton decay comes from dim-9 operator generated via dim-10 $\lambda \mathcal{M}^{-6} R_2 R_2 R_2 \left(H^\dagger\right)^7$ scalar coupling. Resultant, $R_2$ only mediated, proton decay operator is highly suppressed due to $R_2$ LQ mass and the final state phase space.}

The tree level Feynman diagram that contributes to the proton decay, shown in Fig.~\ref{fig:s1_p_decay_2}, is suppressed by the neutrino mass mixing. As a result, all leading proton decay channels in this model involve tiny Dirac neutrino mass. Therefore, proton decay amplitude is proportional to the neutrino mass at leading order and vanishes in the limit of massless neutrinos, which makes proton naturally long lived when neutrino masses are tiny.

The Lagrangian and scalar potential for the model fields content are as follows
\begin{subequations}
\label{eq:lag3}
\begin{align}
    \label{eq:lag_yuk3}
    -\mathcal{L}^{\text{Yuk}}_{\text{BSM}} &= Y_1^{\prime\prime} \Bar{d}\Bar{d} \Bar{S}_{1} + Y_2^{\prime\prime}  \Bar{u}\Bar{\nu} \Bar{S}_{1}^{\dagger} + Y_3^{\prime\prime} Q \Bar{e} R_{2}^\dagger \nonumber \\
 &+ Y_4^{\prime\prime} L \Bar{u} R_2 + Y_5^{\prime\prime} L \Bar{\nu} H 
    + M_\nu \Bar{\nu}\Bar{\nu}  + \text{h.c.} \\
    \label{eq:lag_pot3}
    V &= m_H^2 \left(H^\dagger H\right) + m_{S_1}^2 \left(\Bar{S}_1^{\dagger}\Bar{S}_1\right) + \mu_{R}^2 \left(\Tilde{R}_{2}^\dagger \Tilde{R}_{2}\right) \nonumber \\
    &+ \lambda_H \left(H^\dagger H\right)^2 + \lambda_{S}\left(\Bar{S}_1^{\dagger}\Bar{S}_1\right)^2 + \lambda_{R} \left(\Tilde{R}_{2}^\dagger \Tilde{R}_{2}\right)^2 \nonumber \\
    & + \lambda_{HS} \left(H^\dagger H \right)\left(\Bar{S}_1^{\dagger}\Bar{S}_1\right)+ \lambda_{HR} \left(H^\dagger H \right)\left(R_2^{\dagger}R_2\right)  \nonumber \\
    &+\lambda_{RS}\left(R_2^{\dagger}R_2\right)\left(\Bar{S}_1^{\dagger}\Bar{S}_1\right)
+ \lambda_{HR}^\prime \left(H^\dagger R_2\right)\left(R_2^\dagger H\right)\nonumber \\
    & + \lambda_{HS}^{\prime} \left(H^\dagger \Bar{S}_1\right)\left(\Bar{S}_1^\dagger H\right) + \lambda_{RS}^\prime \left(R_2^\dagger \Bar{S}_1 \right)\left(\Bar{S}_1^\dagger R_2\right) \nonumber \\
&+ \mu_{HRS}\left(H R_2 \Bar{S}_1\right)+\text{h.c.}
\end{align}
\end{subequations}

\begin{figure}[t]
\centering
    \includegraphics[width=0.35\textwidth]{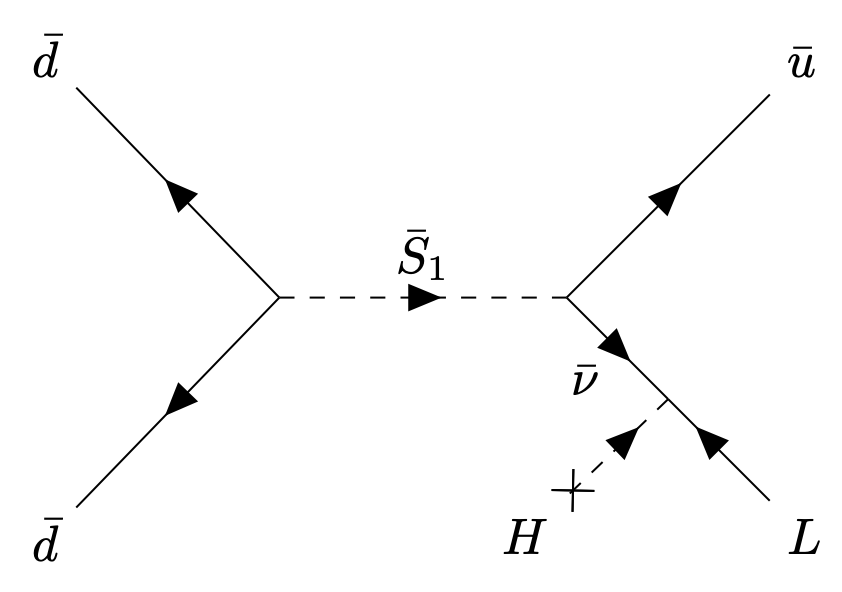}
    \caption{$\Bar{S}_1$ mediated, proportional to Dirac neutrino mass, proton decay Feynman diagram in the framework of neutrino mass mediated proton decay model.}
    \label{fig:s1_p_decay_2}
\end{figure}
The proton decay width is given as 
\begin{widetext}
\begin{subequations}
    \label{eq:s1_p_decay}
    \begin{align}
        \Gamma \left( p\rightarrow e^+ \pi^{0} \right) &= \frac{m_p}{512\pi^3} \left(1-\frac{m_\pi^2}{m_p^2}\right)^2 \left| \left(Y_1^{\prime\prime}\right)^{1a} \frac{m_d^{ab}}{m_{R}} \left(Y_3^{\prime\prime}\right)^{b1} \left(Y_2^{\prime\prime}\right)^{1c} \frac{m_\nu^{cd}}{m_R} \left(Y_4^{\prime\prime}\right)^{d1} \frac{W_0^{l}\left(p\rightarrow e^+ \pi^{0}\right)}{m_R^2} \right. \\
        &\times\left. \left[ H_0 (x_d, x_\nu, x_S, y) + H_1 (x_d, x_\nu, x_S, y) \frac{q^2}{m_R^2} \right] \right|^2 , \nonumber \\
        \Gamma \left( p\rightarrow \nu_\alpha^\dagger \pi^{+} \right) &= \frac{m_p}{32\pi} \left(1-\frac{m_\pi^2}{m_p^2}\right)^2 \left| \left(Y_2^{\prime\prime}\right)^{1\alpha} \left(Y_1^{\prime\prime}\right)^{11} \sqrt{\frac{m_\nu}{M_N}} \frac{W_0^{l}\left(p\rightarrow \nu^\dagger \pi^{+}\right)}{q^2-m_S^2} \right|^2,
    \end{align}
\end{subequations}
\end{widetext}

where $M_N$ is a heavy neutrino Majorana mass, $m_p$ is proton mass, $m_\pi$ is pion mass, $m_\nu$ is light neutrino mass, $q$ is momentum exchanged between quarks in the decay process, and box loop functions ($H_{0,1}$), Kallen $\lambda$, and other relevant definitions are given in the supplementary materials' Sec.~C. $a,b,c$, and $d$ are flavor indices. For proton decay to charged anti-lepton, the final flavor of the anti-lepton is $e^+$, whereas for decay into neutrino, flavor of the later one can be of all three kinds. The proton decay to charged anti-lepton is generated by the Feynman diagram in Fig.~\ref{fig:sce_3_p_decay}, while former's decay into neutrinos is obtained through the Feynman diagram in Fig.~\ref{fig:s1_p_decay_2}.

The proton decay matrix elements and exchanged momentum between the proton's quarks in the decay process are obtained from the lattice calculations~\cite{Aoki:2017puj} and by the means of parity symmetry $W_0^e (p \rightarrow e^+ \pi^0) = \left\langle \pi^0 \right| (ud)_R u_R \left|p\right\rangle = \left\langle \pi^0 \right| (ud)_L u_L \left|p\right\rangle$, $W_0^\nu (p \rightarrow \nu^\dagger \pi^+) = \left\langle \pi^+ \right| (du)_R d_R \left|p\right\rangle = \left\langle \pi^+ \right| (du)_L d_L \left|p\right\rangle$.~\cite{Aoki:2006ib} These values are given in Eq.~\eqref{eq:p_decay_mat_elem}.

Some important facts about this model of proton decay via small neutrino mass include, the compactness and uniqueness of this model. No other leptoquark extension of the SM without BSM symmetries can produce correlation between tiny neutrino masses and proton decay width, hence in this sense, $\Bar{S}_1$ leptoquark is unique. Next, since $\Bar{S}_1$ leptoquark is an EW singlet and couples only to the heavy neutrino SM singlet of all SM leptons, it contributes only to proton decays with final states of charged meson and missing energy ($\Bar{\nu}$). Proton decays via small neutrino masses with final states of charged anti-leptons and neutral mesons can be obtained through one-loop proton decay Feynman diagrams mediated by $R_2$ leptoquark, refer to Fig.~\ref{fig:sce_3_p_decay}. The proton decay of both contributions, tree level $\Bar{S}_1$ and one-loop order $\Bar{S}_1$ and $R_2$ mediated Feynman diagrams' amplitudes are proportional to the neutrino masses.
%
\section{Phenomenological and cosmological constraints}
\label{sec:constraints}
%
%
\subsection{Lepton Flavor Violation}

For the extended scotogenic model (first connection), the Yukawa interaction $Y_4LN\eta^{\dagger}$ generates charged lepton flavor violating (CLFV) processes, $l_i\rightarrow l_j \gamma$.
As studied in ~\cite{Toma:2013zsa,Vicente:2014wga}, the $l_i\rightarrow l_j \gamma$ process currently places the most stringent constraints on the
parameters of the scotogenic neutrino models in a wide area of the parameter space.
This is applicable to the scenario.
The non-observation of the $\mu \rightarrow e \gamma$ leads to $Y_4^{e,\mu}\lesssim 0.01-1$ for mediators with masses of around 1 TeV \cite{Vicente:2014wga}.
For the second connection scenario, those CLFV processes are occurred at three loop level, which are suppressed enough to satisfy the constraints.
Similar to the extended scotogenic model, the Yukawa interactions $LH\bar{\nu}$ in the thrid connection scenario generates the CLFV processes and the coupling constants are constrained by the experiments.
For future prospects of the experimental limits on the CLFV processes are presented in \cite{Calibbi:2017uvl}.
\subsection{Constraints from Search for Leptoquarks (LQs)}
The models we proposed contain scalar LQs as heavy new particles.
As summarized in \cite{ParticleDataGroup:2022pth} , bounds on LQ states are obtained both directly and indirectly.
Direct bounds are derived from their production cross sections at colliders, while indirect bounds are calculated from bounds on LQ-induced two-quark two-lepton interactions, which are obtained from low-energy experiments, or from collider experiments below threshold.
ATLAS and CMS experiments at center of mass energy of 13 TeV lead us to the mass limits on first, second, and third generation LQs which
couple only with the $i$-th generation quarks and the $j$-th generation leptons $(i\neq j)$ without causing conflicts with severe indirect constraints.
The systematic search strategies as well as the results are presented in detail in \cite{ParticleDataGroup:2022pth, Diaz:2017lit, Schmaltz:2018nls}.
On the other hand, their couplings can be used to resolve the anomalies in rare $B$ meson decays and the anomalous magnetic moment of the muon \cite{ Schmaltz:2018nls,Muller:2018nwq}.
The recent analysis for the searches of $lljj$ final states at the LHC produced via pair production of scalar leptoquarks shows that their masses below around 1.7 TeV are excluded by the LHC data \cite{Bhaskar:2023ftn}.
\subsection{Dark Matter}
As mentioned before, the extended scotogenic model has a DM candidate. The lightest $N$ is supposedly DM and was thermally produced and frozen out in the early Universe. 
The annihilation processes, $NN\rightarrow LL$ mediated by $\eta$, can contribute to the relic density of DM. For those processes to be efficient enough, $Y_4$ should be large, which leads to a mild tension between the relic abundance of DM and upper
bounds obtained from CLFV processes discussed above \cite{Calibbi:2017uvl}.
There exist another contributions to the relic density of DM, which stem from
the annihilation processes, $NN\rightarrow QQ$ mediated by $R_{2D}$.
Not so small values of $Y_2$ is required to satisfy the right amount of the relic density without any tension.
%
\section{Discussion}
\label{sec:discussion}
The connection between the origins of neutrino mass and proton decay establishes a link between lepton and quark sectors of the SM, which requires BSM physics testable at future colliders like FCC, etc. Bridging the gap between quark and lepton sectors naturally calls for leptoquarks, which appear in all three models outlined in the previous sections. Furthermore, the presence of leptoquarks will certainly inspire flavor physicists to resolve the flavor physics anomalies within the context of models connecting neutrino mass and proton decay, and possibly dark matter.

Another important follow up to be digged into is unification. Unification naturally predicts proton decay, which serves as a smoking gun signature of the GUT framework. Unifying the proton decay and neutrino mass origins within the study of GUT framework is no easy, nevertheless a very charming, path which requires gluing the proton decay mechanism to the neutrino mass generation in the context of a larger symmetries. This will induce the new class of GUT models, distinct from the current most investigated GUT model like $SU(5), SO(10), E_6$, etc.

The connection between neutrino mass and proton decay origins is capable of bringing new look into several directions of high energy physics.
%
\section{Conclusion}
\label{sec:conc}
After outlining the three sample models for realizing the correlation between proton lifetime longevity and smallness of the neutrino mass, brief discussion of the idea and future prospective directions was given. We conclude that connection between neutrino mass and proton decay origins is possible and attractive concept. This framework gives opportunity to link the open questions of neutrino mass and proton decay to each other, and possibly to the existence of dark matter. This way justifying the smallness of neutrino masses will explain the longevity of the proton lifetime and vice versa. This framework also complements the proton decay experimental searches with the neutrino search data and possibly dark matter search experimental data. We presented only the most minimalistic realizations of the idea for the three causal connections between neutrino mass and proton lifetime longevity. The concept of correlation between neutrino mass and proton decay origin will bring new ideas into the two open questions and beyond.
%
\acknowledgments
The work was supported by the National Natural Science Fund of China Grant No.~12350410373 (O.~P.) and by the National Research Foundation of Korea under grant NRF-2023R1A2C100609111 (S.~K.). All Feynman diagrams were created using \small{TikZ-Feynman LateX} package~\cite{Ellis:2016jkw}.

The order of authors' names is alphabetical.
%
%
%
%
\bibliography{references}
\setcounter{figure}{0}
\begin{widetext}
%
\preprint{2412.xxxx}
\title{\boldmath \color{BrickRed}Pathways to proton's stability via naturally small neutrino masses \\ ~ \\ {\normalfont\color{black}\emph{Supplementary Materials}}}
%
\author{Sin Kyu Kang \orcidlink{0000-0001-7508-3881}}
\author{Oleg Popov \orcidlink{0000-0002-0249-8493}}
%
%
%
%
\maketitle 
\begin{center}
\Large{\boldmath \color{BrickRed}Pathways to proton's stability via naturally small neutrino masses \\ ~ \\ {\normalfont\color{black}\emph{Supplementary Materials}}}
\end{center}
\appendix
All three connection types between neutrino mass and proton decay have their advantages and require distinct minimal BSM content. The concept of connection between neutrino mass and proton decay is schematically represented in Fig.~\ref{fig:concept_scheme}.

\begin{figure}[!h]
    \centering
    \begin{subfigure}[b]{0.3\textwidth}
        \includegraphics[width=0.9\textwidth]{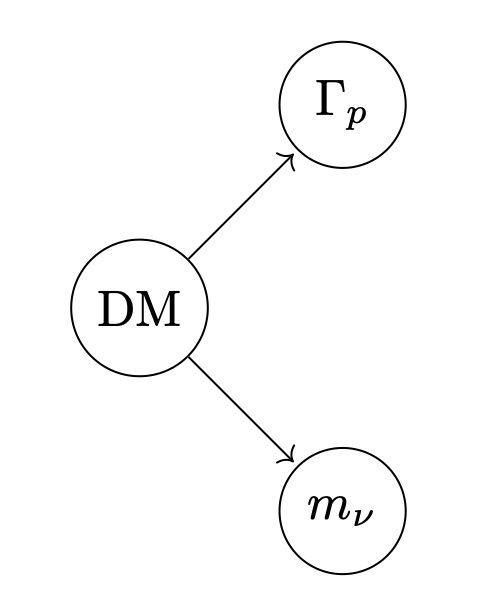}
        \caption{Schematic representation of the generation of the neutrino mass and proton decay from common origin.}
        \label{fig:connection_1}
    \end{subfigure}
    \begin{subfigure}[b]{0.33\textwidth}
        \includegraphics[width=0.9\textwidth]{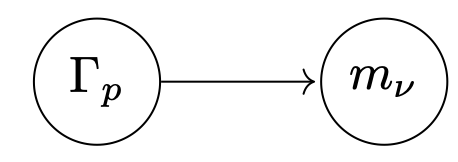}
        \caption{Schematic representation of the generation of the neutrino mass from proton decay origin.}
        \label{fig:connection_2}
    \end{subfigure}
    \begin{subfigure}[b]{0.33\textwidth}
        \includegraphics[width=0.9\textwidth]{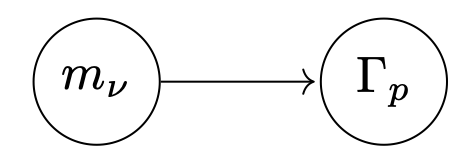}
        \caption{Schematic representation of the generation of the proton decay from neutrino mass origin.}
        \label{fig:connection_3}
    \end{subfigure}
    \caption{Schematic representations of the concept of connection between neutrino mass and proton decay.}
    \label{fig:concept_scheme}
\end{figure}
\section{Extended scotogenic model}
This section provides the supplementary tables, figures, and equations for the extended scotogenic model.

Tab.~\ref{tab:sce_1_fields} enlists the field content of the extended scotogenic model.

\begin{table}[h]
    \centering
    \begin{tabular}{ccccc}
        \hline \hline
        Fields & $SU(3)_c$ & $SU(2)_L$ & $U(1)_Y$ & $\mathbb{Z}_2$ \\ \hline
        $Q$ & $\pmb{3}$ & $\pmb{2}$ & $~\frac{1}{6}$ & $+$ \\
        $\Bar{u}$ & $\pmb{\Bar{3}}$ & $\pmb{1}$ & $-\frac{2}{3}$ & $+$ \\
        $\Bar{d}$ & $\pmb{\Bar{3}}$ & $\pmb{1}$ & $~\frac{1}{3}$ & $+$ \\
        $L$ & $\pmb{1}$ & $\pmb{2}$ & $-\frac{1}{2}$ & $+$ \\
        $\Bar{e}$ & $\pmb{1}$ & $\pmb{1}$ & $1$ & $+$ \\
        $N$ & $\pmb{1}$ & $\pmb{1}$ & $0$ & $-$ \\
        $D,\Bar{D}^\dagger$ & $\pmb{3}$ & $\pmb{1}$ & $-\frac{1}{3}$ & $-$ \\ \hline
        $H$ & $\pmb{1}$ & $\pmb{2}$ & $~\frac{1}{2}$ & $+$ \\
        $\Tilde{R}_{2D}$ & $\pmb{3}$ & $\pmb{2}$ & $~\frac{1}{6}$ & $-$ \\
        $\eta$ & $\pmb{1}$ & $\pmb{2}$ & $-\frac{1}{2}$ & $-$ \\ \hline \hline
    \end{tabular}
    \caption{Field content of extended scotogenic model.}
    \label{tab:sce_1_fields}
\end{table}

Small neutrino masses in the extended scotogenic model are generated by the canonical scotogenic neutrino mass, the Feynman diagram of which is given in Fig.~\ref{fig:sce_1_scoto_mnu_2006}.

\begin{figure}[ht]
\centering
\includegraphics[width=0.5\textwidth]{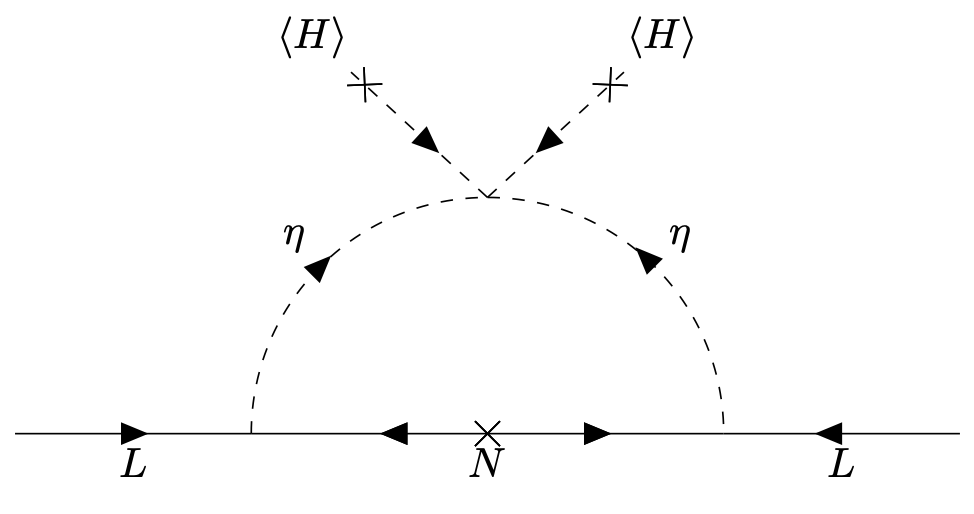}
\caption{Scotogenic neutrino mass.}
\label{fig:sce_1_scoto_mnu_2006}
\end{figure}

Box-loop integral functions relevant for the radiative proton decay of Fig.~{\color{red}1} of the main manuscript are given as follows:
%
%
\begin{subequations}
    \label{eq:fg_functions}
    \begin{align}
        &F_0 (x_N, x_D, y) = \left\{ (1+x_D) (x_N-1) y \ln x_D - (x_D-1)\left[(x_D-x_N) (x_N-1)\ln\left(\frac{x_D}{x_N}\right) - (1+x_N) y \ln x_N \right] \right. \\
        &\left.- (x_D-1)(x_N-1) \lambda^{1/2}(x_D,x_N,y) \left[ \ln(4 x_D x_N) -2 \ln\left(x_D + x_N - y + \lambda^{1/2}(x_D,x_N,y) \right) \right] \right\} \nonumber \\
        &\times \left[ 2 y (x_D-1)(x_N-1) \left((x_D-1)(x_N-1) + y\right) \right]^{-1}, \nonumber \\
        &F_1 (x_N, x_D, y) = \left\{ 4 y \left[ (x_D - x_N)^2 + y \frac{1 - 6 x_D x_N + 2 x_N^2 + 2 x_D^2 + x_D^2 x_N^2}{(x_D-1)(x_N-1)} + y^2 \frac{1 - 4x_D x_N + x_D^2 + x_N^2 + x_D^2 x_N^2}{(x_D-1)^2(x_N-1)^2} \right] \right. \nonumber \\
        &+\left( 2 \left[(x_D-1)^2(x_D-x_N)^2 - 2 y (x_D-1)\left(-x_N + x_D (-2 + x_D + 2 x_N) \right) + y^2 \left( 1 - 4 x_D + x_D^2 \right) \right] \right. \nonumber \\
        &\left.\times \left[ -x_N + y + x_D (1 - x_D + x_N + y) \right] \ln x_D (x_D-1)^{-3} + \left( x_D \leftrightarrow x_N \right) \right) \nonumber \\
        &\left.- 2 \lambda^{3/2}(x_D,x_N,y) \left[ \ln(4 x_D x_N) - 2 \ln \left( x_D + x_N - y + \lambda^{1/2}(x_D,x_N,y) \right) \right] \right\} \frac{
        1}{24 y^2 \left[ (x_D-1)(x_N-1) + y \right]^2}, \\
        \label{eq:lambda_func_1}
        &\hspace{4cm} \text{where } \lambda\left(a,b,c\right) = a^2 + b^2 + c^2 - 2 a b - 2 a c - 2 b c, \\
        \label{eq:x_set_def_1}
        &\hspace{4.5cm} \text{and } x_N = \frac{M_N^2}{m_R^2}, \quad x_D = \frac{M_D^2}{m_R^2}, \quad y = \frac{m_p^2}{m_R^2}.
    \end{align}
\end{subequations}
%

Here and hereafter, all relevant quantities are defined in dimensionless units for the convenience of the reader.
%
\FloatBarrier
\section{Proton decay induced neutrino mass origin}

This section provides the supplementary tables and figures for the scenario, where neutrino masses are generated by the means of proton decay.  Which in turn, leads to naturally small neutrino masses for the case of long proton lifetime.

The field content of this model is given in Tab.~\ref{tab:sce_2_fields}, where the SM was extended with $U(1)_{F}$ global symmetry and two copies of $S_3$ leptoquark with two different F charges. F charge is defined as $F=3B-L$. This is done in order to break $U(1)_F$ symmetry softly by dim-2 term $S_3^\dagger S_3^\prime + $ h.c.
\begin{table}[h]
    \centering
    \begin{tabular}{ccccc}
        \hline \hline
        Fields & $SU(3)_c$ & $SU(2)_L$ & $U(1)_Y$ & $U(1)_F$ \\ \hline
        $Q$ & $\pmb{3}$ & $\pmb{2}$ & $~\frac{1}{6}$ & $~~1$ \\
        $\Bar{u}$ & $\pmb{\Bar{3}}$ & $\pmb{1}$ & $-\frac{2}{3}$ & $-1$ \\
        $\Bar{d}$ & $\pmb{\Bar{3}}$ & $\pmb{1}$ & $~\frac{1}{3}$ & $-1$ \\
        $L$ & $\pmb{1}$ & $\pmb{2}$ & $-\frac{1}{2}$ & $-1$ \\
        $\Bar{e}$ & $\pmb{1}$ & $\pmb{1}$ & $1$ & $~~1$ \\ \hline
        $H$ & $\pmb{1}$ & $\pmb{2}$ & $~\frac{1}{2}$ & $~~0$ \\
        $S_3,S_3^\prime$ & $\pmb{\Bar{3}}$ & $\pmb{3}$ & $~\frac{1}{3}$ & $0,2$ \\ \hline \hline
    \end{tabular}
    \caption{Field content of neutrino mass via proton stability model, where $F=3B-L$.}
    \label{tab:sce_2_fields}
\end{table}

The Feynman diagram that contributes to the two-body proton decay and is proportional to the $S_3$ and $S_3^\prime$ mixing is given in Fig.~\ref{fig:p_decay_2b_df_sce_2}. The effective neutrino mass that is generated in the present scenario is illustrated in Fig.~\ref{fig:3l_weinberg_eft_sce_2}. In the case of proton decay width approaching zero the neutrino masses become naturally small.

\begin{figure}[h]
\centering
    \begin{subfigure}[t]{0.45\textwidth}
    \centering
    \includegraphics[width=0.9\textwidth]{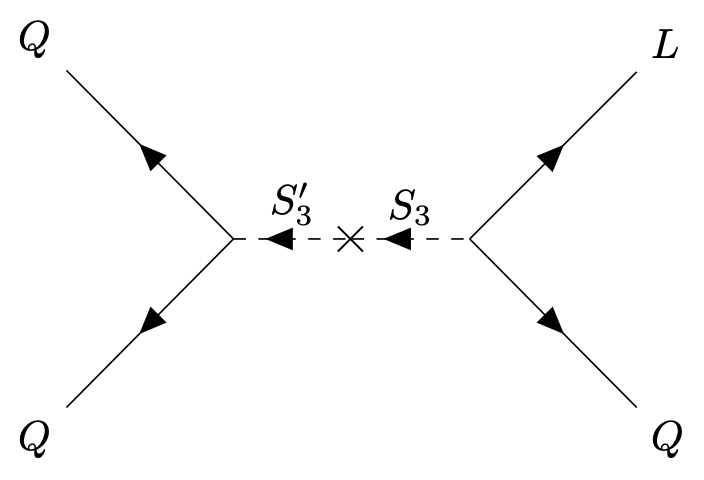}
    \caption{Feynman diagram contributing to two(three) body proton decay ($p\rightarrow \pi^+\nu, \pi^+\pi^+ e^-$) through $U(1)_F$ soft breaking.}
    \label{fig:p_decay_2b_df_sce_2}
    \end{subfigure}
    \begin{subfigure}[t]{0.45\textwidth}
    \centering
    \includegraphics[width=0.8\textwidth]{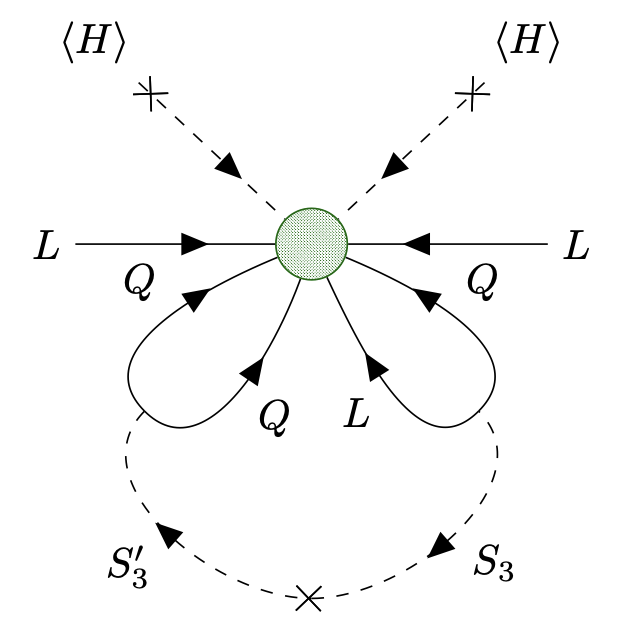}
    \caption{Neutrino mass 3-loop Weinberg operator generated through the proton decay operator.}
    \label{fig:3l_weinberg_eft_sce_2}
    \end{subfigure}
    \caption{Feynman diagrams that contribute to proton decay induced neutrino mass origin framework.}
    \label{fig:sce_2_dia_1}
\end{figure}
%
%
\FloatBarrier
\section{Neutrino mass mediated proton decay}

The field content of the minimal model for this scenario is given in Tab.~\ref{tab:sce_3_fields}.

\begin{table}[h]
    \centering
    \begin{tabular}{cccc}
        \hline \hline
        Fields & $SU(3)_c$ & $SU(2)_L$ & $U(1)_Y$ \\ \hline
        $Q$ & $\pmb{3}$ & $\pmb{2}$ & $~\frac{1}{6}$ \\
        $\Bar{u}$ & $\pmb{\Bar{3}}$ & $\pmb{1}$ & $-\frac{2}{3}$ \\
        $\Bar{d}$ & $\pmb{\Bar{3}}$ & $\pmb{1}$ & $~\frac{1}{3}$ \\
        $L$ & $\pmb{1}$ & $\pmb{2}$ & $-\frac{1}{2}$ \\
        $\Bar{e}$ & $\pmb{1}$ & $\pmb{1}$ & $1$ \\
        $\Bar{\nu}$ & $\pmb{1}$ & $\pmb{1}$ & $0$ \\ \hline
        $H$ & $\pmb{1}$ & $\pmb{2}$ & $~\frac{1}{2}$ \\
        $R_2$ & $\pmb{3}$ & $\pmb{2}$ & $~\frac{7}{6}$ \\
        $\Bar{S}_1$ & $\pmb{\Bar{3}}$ & $\pmb{1}$ & $-\frac{2}{3}$ \\ \hline \hline
    \end{tabular}
    \caption{
    Field content of proton stability via smallness of neutrino mass model.}
    \label{tab:sce_3_fields}
\end{table}

Box-loop integral functions relevant for the radiative proton decay of Fig.~{\color{red}3} of the main manuscript are given as follows:

\begin{subequations}
    \label{eq:box_h_loop_func}
    \begin{align}
        \label{eq:h0_func}
        y \left( x_S - 1 \right) H_0 \left( x_d, x_\nu, y \right) &= \left[ \left\{ f_{-}(x_d-x_S) - f_{-}(x_d-x_S-y) + f_{+}(x_d-x_S) - f_{+}(x_d-x_S-y) \right.\right. \\ 
        &\left.\left.- g(x_d-x_S) + g(x_d-x_S-y) \right\} - \left( x_S \rightarrow 1 \right) \right], \nonumber \\
        \label{eq:f_func}
        f_{\pm}(x) &= \text{DiLog}\left[ \frac{ 2 x }{x_d - 2 x_S + x_\nu  - y \pm \sqrt{\lambda \left( x_d, x_S, y \right)}}, \mp y x \right], \\
        \label{eq:g_func}
        g(x) &= \text{PolyLog} \left[ 2, \frac{ x \left( x_\nu - x_S \right) }{ \left( x_d - x_S \right) \left( x_\nu - x_S \right) + y x_S } \right], \\
        \label{eq:h1_func}
        y^2 \left( x_S - 1 \right)^3 H_1 \left( x_d, x_\nu, y \right) &= \left[ (x_S - 1) y + \frac{ \left(x_S - x_d\right) y \ln x_d}{x_d - 1} + \frac{\left( x_d - 1 \right) x_S \ln \left( \frac{x_d}{x_S} \right)}{x_d - x_S} \right. \\
        &\left.- \frac{1}{2} \left( x_S - 1 \right) \left( x_d - x_\nu \right) \ln \left( \frac{x_d}{x_\nu} \right) - \left( x_S - 1 \right) \lambda^{1/2} \left( x_d, x_\nu, y \right) \left\{ \ln \left( 2 x_d \right) \right. \right. \nonumber \\
        &\left.\left.- \ln \left[ x_d + x_\nu - y + \lambda^{1/2} \left( x_d, x_\nu, y \right) \right] \right\} + \left\{ x_d \left( 1 + x_S - x_\nu \right) - x_S - \frac{y}{2} \left( x_S + 1 \right) \right\} \right. \nonumber \\
        &\left. \times y H_0 \left( x_d, x_\nu, y \right) \left( x_S - 1 \right) + \left( x_d \longleftrightarrow x_\nu \right) \right], \nonumber \\
        \label{eq:lambda_func_2}
        \hspace{0cm} \text{where } \lambda\left(a,b,c\right) &= a^2 + b^2 + c^2 - 2 a b - 2 a c - 2 b c, \\
        \label{eq:x_set_2}
        \hspace{0cm} \text{and } x_d &= \frac{m_d^2}{m_R^2}, \quad x_\nu = \frac{m_\nu^2}{m_R^2}, \quad x_S = \frac{m_S^2}{m_R^2}, \quad y = \frac{m_p^2}{m_R^2}.
    \end{align}
\end{subequations}

In order to avoid proton decay that is independent of the neutrino mass, SM is extended by the $\Bar{S}_1$ leptoquark and neutrinos are taken to be Majorana, with seesaw scale heavy sterile neutrino component. The relevant kinematically forbidden Feynman diagram is given in Fig.~\ref{fig:s1_p_decay_1}. 

\begin{figure}
\centering
    \includegraphics[width=0.4\textwidth]{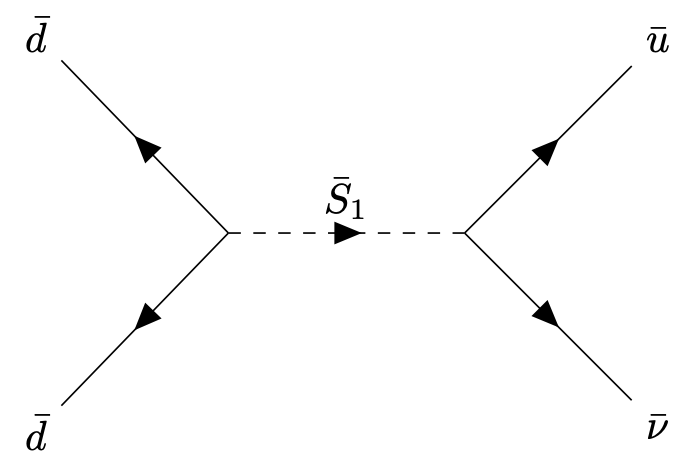}
    \caption{$\Bar{S}_1$ mediated proton decay feynman diagram which is kinematically forbidden and that otherwise would contribute to proton decay width.}
    \label{fig:s1_p_decay_1}
\end{figure}

Seesaw-I mechanism is given in Fig.~\ref{fig:sce_3_dia_1}.

\begin{figure}[h]
\centering
    \includegraphics[width=0.4\textwidth]{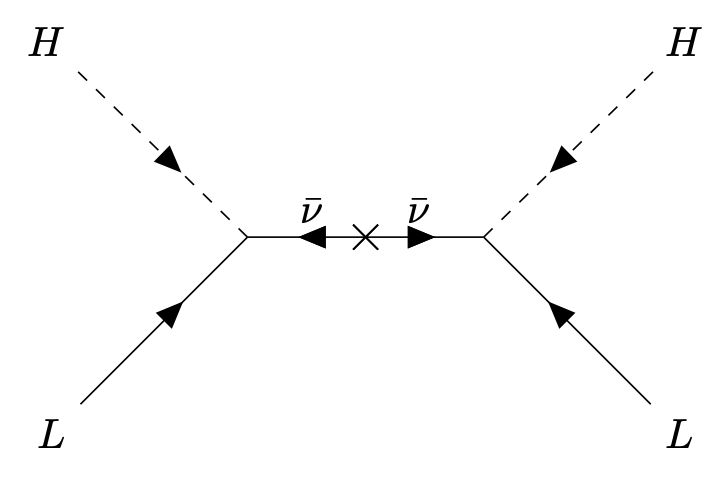}
    \caption{Seesaw-I mechanism for the neutrino mass mediated proton decay.}
    \label{fig:sce_3_dia_1}
\end{figure}

%
\section{Supplementary Mathematical Functions}
\label{sec:sup_math_func}
The aforementioned mathematical functions used in the calculations can be approximated as follows.

\begin{subequations}
    \label{eq:sup_math_func}
    \begin{align}
        \text{DiLog} \left[ x, \mathbb{R}^{\pm} \right] &= 
            \begin{cases}
                \frac{\pi^2}{6} + (x-1)\left[ 1 - \ln (x-1) \pm \imath \pi \right] & \quad x \gtrsim 1 \\
                \frac{\pi^2}{6} - (1-x)\left[ 1 - \ln (1-x) \right] & \quad x \lesssim 1
            \end{cases}, \\
        \text{PolyLog} \left[ 2, x \right] &= 
            \begin{cases}
                \frac{\pi^2}{6} + (x-1)\left[ 1 - \ln (x-1) - \imath \pi \right] & \quad x \gtrsim 1 \\
                \frac{\pi^2}{6} - (1-x)\left[ 1 - \ln (1-x) \right] & \quad x \lesssim 1
            \end{cases},
    \end{align}
\end{subequations}

where $\mathbb{R}^{\pm}$ stands for positive and negative real numbers.
%

%
%
%
%

%

\end{widetext}
\end{document}